\documentclass[lettersize,journal]{IEEEtran}
\usepackage{mathptmx} 
\usepackage{footnote}
\makesavenoteenv{tabular}
\makesavenoteenv{table}
\usepackage{pgfplots}

\usepackage{bbm}
\usepackage{booktabs}
\usepackage{xcolor}
\usepackage[hidelinks, hyperfootnotes=true]{hyperref}
\usepackage{nomencl}
\usepgfplotslibrary{fillbetween}
\definecolor{lightgreen}{rgb}{0.56, 0.93, 0.56} 

\usepackage{amsmath,amsfonts}
\usepackage{algorithmic}
\usepackage{algorithm}
\usepackage{array}
\usepackage[caption=false,font=normalsize,labelfont=sf,textfont=sf]{subfig}
\usepackage{textcomp}
\usepackage{stfloats}
\usepackage{url}
\usepackage{verbatim}
\usepackage{graphicx}
\usepackage{cite}
\usepackage{color, colortbl} 
\usepackage{etoolbox}
\usepackage{nomencl}
\makenomenclature
\renewcommand\nomgroup[1]{%
  \item[\itshape
  \ifstrequal{#1}{R}{Residential Load Profiles:}{%
  \ifstrequal{#1}{F}{Full Convolutional Profile Flow:}{%
  \ifstrequal{#1}{E}{Evaluation Metrics:}{}}}%
]}

\begin{document}


\title{Foundation Twins: A New Generation of Power Systems Digital Twins using Foundation AI Models}

\author{Pedro P. Vergara,~\IEEEmembership{Senior Member,~IEEE}
\thanks{Pedro P. Vergara is with the Intelligent Electrical Power Grids (IEPG) Group, Delft University of Technology, 2628 CD Delft, The Netherlands (e-mail: P.P.VergaraBarrios@tudelft.nl).}

}


\maketitle

\begin{abstract}
Power systems are inherently multi-timescale systems, with different physical phenomena and decision-making processes spanning multiple timescales, time horizons, and geographic scopes. I envision power systems digital twins (DTs) as powerful modeling and simulation tools that can accelerate and improve decision-making across different time scales and geographic scopes. However, until now, research has not delivered such a vision, and power systems DTs remain a concept distant from implementation. This is not a regular research paper. This is a position paper that outlines my vision for developing a new generation of power systems DTs that leverage recent advances in artificial intelligence (AI) and machine learning (ML). I call these \textit{Foundation Twins}. Foundation Twins combines the generalization features of foundation models with the decision-making capabilities of reinforcement learning (RL) architectures to deliver the envisioned power systems DTs. 
\end{abstract} 

\section{Introduction}\label{intro}
\IEEEPARstart{D}{igital} Twins (DTs), in the context of power systems, can be defined as a ...\textit{"collection of modules and models (based on multi-physics simulation) integrated into a single (software) ecosystem ... aimed at mirroring the real-time operation of the power systems and supporting its long-term planning."}\cite{Zomerdijk2024}. According to this definition, DTs are expected to be composed of several (multi-physics) simulation models that serve as the basis for other processes. Some such processes may include dispatching distributed energy resources (DERs), frequency and voltage control, and managing transformer and cable capacity. In general, these processes operate at different timescales, horizons, and geographic scopes, suggesting that DTs must provide multi-timescale and multi-geographic decision-making capabilities. Until now, such capabilities have been missing, and power systems DTs remain a concept distant from deployment. The need for multi-timescale decision-making is readily apparent from a practical perspective. For instance, a single DT that coordinates across multiple timescales could help dispatch DERs in the market (slow dynamics) while accounting for power system stability (fast dynamics), thereby avoiding, for example, voltage and frequency issues. As a result, by accounting simultaneously for fast and slow dynamics, DTs can bridge the current gap between power system operation and planning frameworks, thereby increasing its resilience (e.g., against blackouts).

\begin{figure}[ht]
  \centering
  \includegraphics[width=0.95\columnwidth]{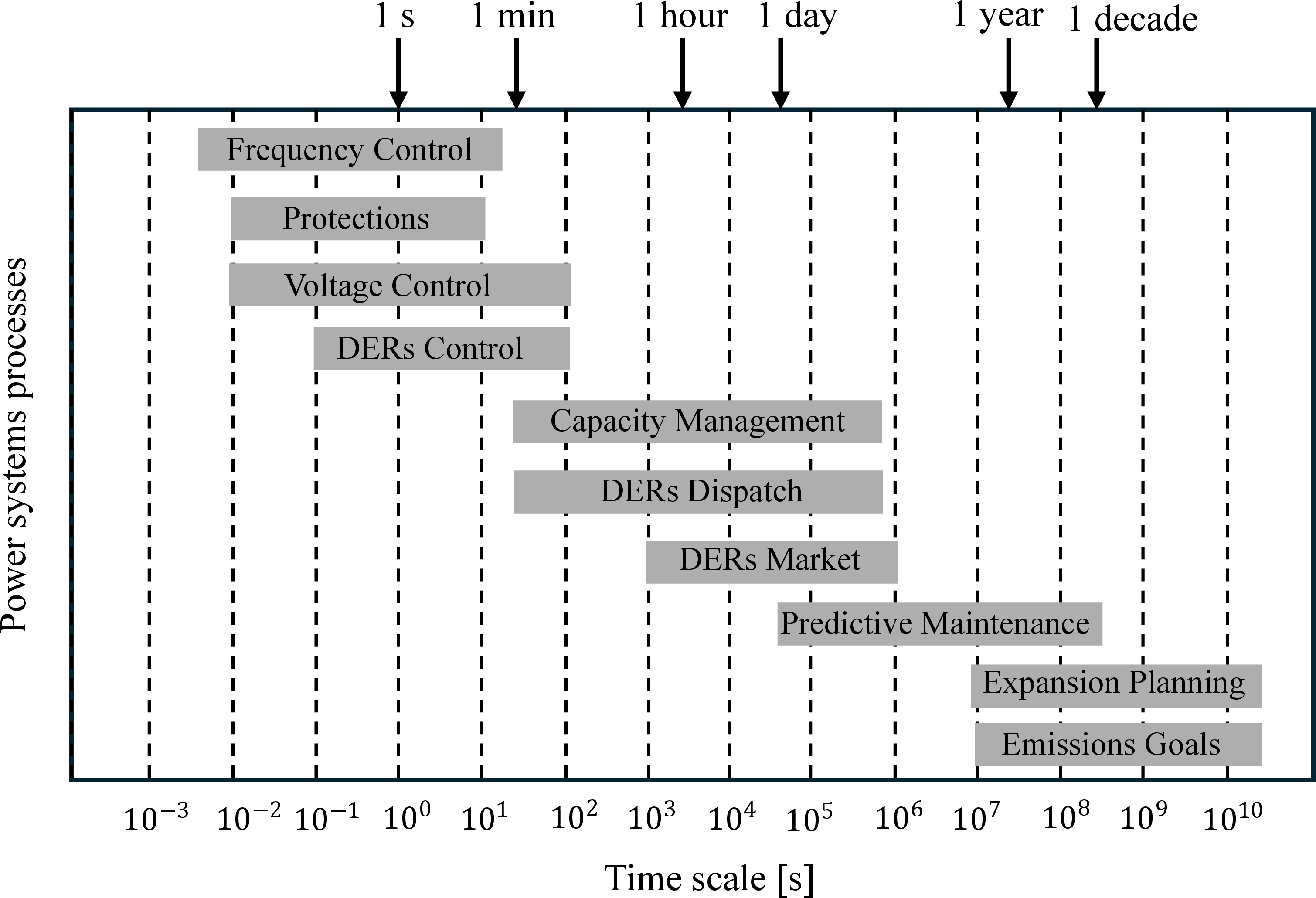}
  \caption{Approximated power systems processes and their different timescales. Taking into account all power systems processes in a unique model would result in an untractable model.}
  \label{fig:power_system_scales}
\end{figure}

\subsection{Multi-Timescale Power Systems}\label{sec:multi-time-scale-ps}
Power systems are composed of a set of devices (i.e., assets, resources) that can be described by continuous dynamics and discrete events. A simple example, synchronous machines, can be modeled using a set of non-linear ordinary differential equations (ODEs) with proper parameters. Other devices, such as controllers, can also be modeled using ODEs. In discrete events, Boolean or discrete variables can be used to model, for instance, switch connection statuses or tap values of tap-changing transformers.  

In power systems analysis, the time constants of different physical phenomena, as well as some decision-making processes, can span a wide range of different timescales and horizons. Fig.~\ref{fig:power_system_scales} presents the time scales of a variety of typical power system decision-making processes. Since the timescale spans from microseconds to decades, taking into account the dynamics of all phenomena and processes in a unique model would result in an untractable approach.

A common solution to this issue is to split the general state vector of system variables $s$ into three sub-vectors~\cite{Milano2010}: (i) state variables $s_{I}$ characterized by slow dynamics, (ii) state variables $s_{II}$ characterized by dynamics of interest, and (iii) state variables $s_{III}$ characterized by fast dynamics. The consequence of this division is that, from the perspective of the state variables of interest $s_{II}$, the time evolution of state variables $s_{I}$ can be considered so slow, while the time evolution of variables $s_{III}$ can be considered so fast, that their variations can be disregarded. As a result, continuous dynamics can now be modeled using nonlinear differential-algebraic equations (DAEs) using both continuous and discrete variables. Although this modeling approach enables separate models for different physical phenomena and decision-making processes, the multi-timescale nature of power system infrastructure persists. Models that enable proper transfer of (fast and slow) dynamics, information, uncertainty, errors, and decisions across timescales are urgently needed. The question that remains is: \textit {How can we develop DTs that provide decision-making capabilities across multiple timescales, horizons, and geographic scopes, without relying on separate physics-based models that exploit dynamics separation?}. I believe that recent developments in artificial intelligence (AI) and machine learning (ML), particularly in foundation models, complemented by the decision-making capabilities of reinforcement learning (RL), can help DTs deliver these expected features. 

\begin{figure*}[ht]
  \centering
  \includegraphics[width=\linewidth]{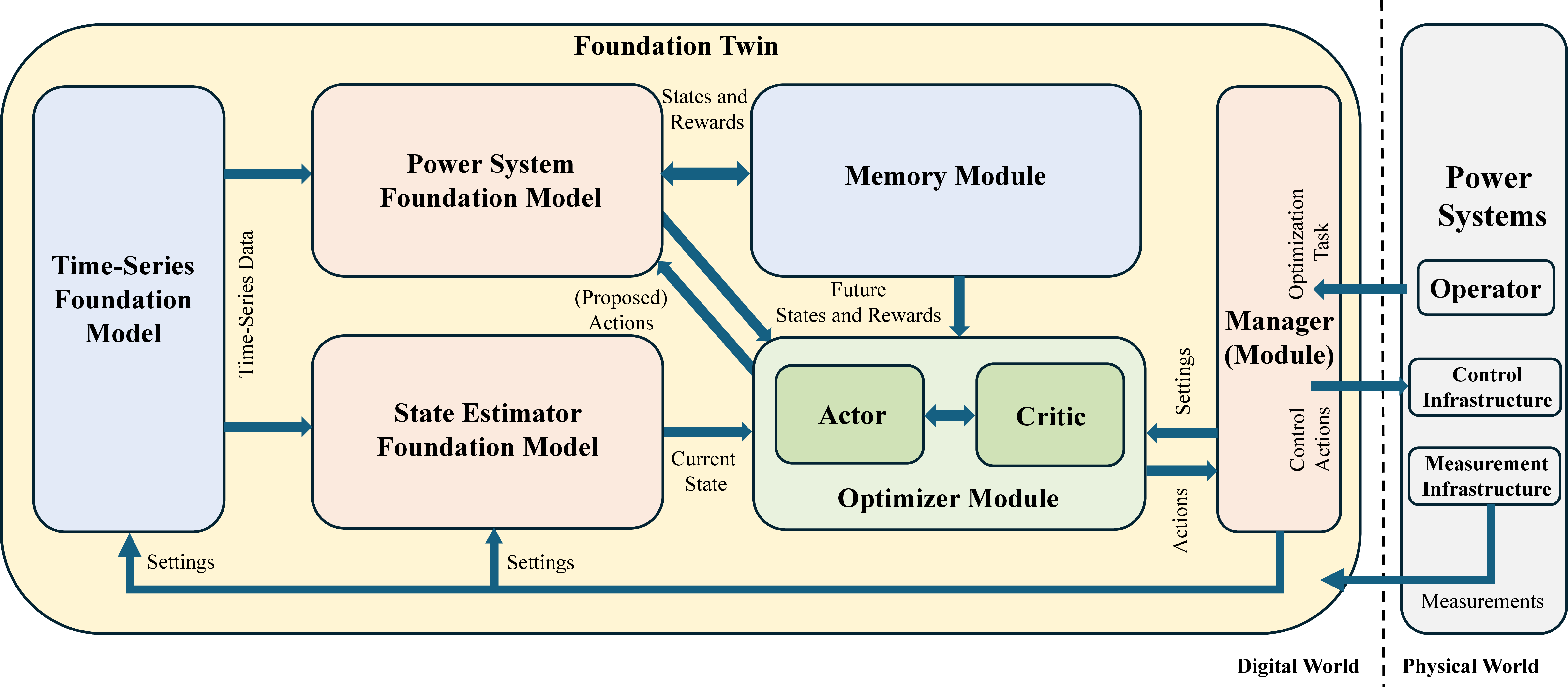}
  \caption{Foundation Twins architecture composed of several foundation AI models to enhance simulation capabilities, complemented by an RL architecture for decision-making. The \textit{Manager} is responsible for coordinating all other models and modules, as well as for interfacing measurements and control actions with the power systems infrastructure. Measurements from the physical power system are fed into the \textit{Time-Series Foundation Model} at different time resolutions. The \textit{Power Systems Foundation Model} serves as a simulator, while the \textit{State Estimator Foundation Model} uses time-series data to estimate the current system state. Based on the current state, the \textit{Optimizer Module} proposes future actions using an \textit{Actor}, while expected (accumulated) rewards are estimated via the \textit{Critic}. Future system states are provided via the \textit{Power System Foundation Model}. Finally, the \textit{Memory Model} stores state-action transitions (and their rewards) for future reference by the Optimizer Module.}
  \label{fig:architecture}
\end{figure*}

\subsection{Foundation Models}
Foundation models have recently emerged as a new learning paradigm in AI. These models learn from large datasets through self-supervision and have proved to generalize across many applications. Unlike supervised learning models, foundation models can generalize across multiple learning tasks with a single architecture, often based on a transformer architecture with an encoder and decoder components. These components are important because they enable the foundation model to reconstruct masked parts of the input data, thereby allowing it to learn from unannotated data. Successful examples of foundation models include the now-commercial large language models (LLMs) such as ChatGPT and LeChat, but their applications extend to time-series data, audio, and video. The potential of foundation AI models for power systems has already been discussed in~\cite{Hamann2024}, where features such as the ability to generalize across multiple tasks with a single architecture and to learn from multimodal data sources have been highlighted as crucial for power systems applications. I argue that although these aspects are important, crucial features required for power system applications are still missing (which I discuss here as well). Moreover, I also argue that a single (and large) model may not be sufficient to deploy the expected capabilities. Instead, I foresee the use of multiple specialized foundation models, properly orchestrated to enhance multi-timescale simulation capabilities, complemented by an RL architecture to deploy decision-making. This is the general idea behind the concept that I introduce in this paper.  

\subsection{Contributions}
This is not a regular research paper. This paper aims to outline a vision for developing a new generation of power systems DTs using recent advances in foundation AI models and RL. Many of the ideas described in this paper have been formulated by various authors in different contexts, and its primary contribution is to assemble them into a consistent whole, proposing a new concept that I call \textit{Foundation Twins}. My expectation is that this new concept (and this paper) will guide future research works to provide the missing multi-timescale decision-making capabilities of power systems DTs.

\section{Foundation Twins}\label{sec:foundational-twins}

Fig.~\ref{fig:architecture} presents the architecture for the proposed \textit{Foundation Twins}, inspired somewhat from the intelligence architecture proposed in~\cite{LeCun2022}. Foundation Twins are composed of a set of foundation AI models with clearly defined boundaries, tasks, and capabilities to enhance multi-timescale simulation, complemented by an RL architecture for decision-making. In this section, I will dive into the details of each foundation model's (and modules') capabilities, as well as the challenges foreseen for their development and implementation. 

\subsection{Architecture Execution}
The \textbf{Manager (Module)} is responsible for taking the main optimization task from the power systems operator (i.e., via a proper interface) and configuring other foundation models and modules to achieve the objective defined by this optimization task. In particular, the \textit{Manager} directly communicates with the \textit{State Estimator Foundation Model}, \textit{Time-Series Foundation Model}, and the \textit{Optimizer Module}, defining the proper settings for their execution. 

The \textbf{State Estimator Foundation Model} main function is to estimate the power system's current state at the appropriate timescale as defined by the \textit{Manager}. To provide an estimate of the current state, the \textit{State Estimator Foundation Model} uses time-series data provided by the \textbf{Time-Series Foundation Model}. Thus, the main function of the \textit{Time-Series Foundation Model} is to provide the appropriate data at the appropriate time scale. For a given optimization task, not all measurement data and/or time scales may be relevant or available for estimating the current power system state (see Sec.\ref{sec:time-series-foundation}). 

Using the current system state information, the \textbf{Optimizer Module} is executed. The main function of this module is to execute the appropriate training and/or deployment procedure according to a predefined policy (see Sec.~\ref{sec:optimizer}). To do this, the \textit{Optimizer Module} deploys an \textbf{Actor Model} and \textbf{Critic Model}, following the same architecture as in actor-critic RL algorithms. The \textit{Actor} is responsible for computing action proposals for the current state, while the \textit{Critic} produces estimations of future (accumulated) rewards, based on reward estimates provided by the \textit{Power System Foundation Model}.  In general, the \textit{Critic} and the \textit{Power System Foundation Model} are responsible for estimating the possible outcomes of any (current or future) action proposed by the \textit{Actor}. 

The \textbf{Power System Foundation Model} is the most important foundation model of the proposed architecture. Its main functionality is to predict future states of the power system using advanced multi-timescale physics simulation capabilities; thus, it can be seen as a high-fidelity simulator of the physical power system (see Sec.~\ref{sec:power-syste-foundation}). The \textit{Power System Foundation Model} predicts the natural evolution of the power system across different timescales and horizons, driven by actions proposed by the \textit{Actor}. Future states would need to be complemented by using future realizations (or forecasts) of (some) power system variables (e.g., renewable generation and demand), provided by the \textit{Time-Series Foundation Model}. Simultaneously, the \textit{Power System Foundation Model} provides an estimate of future rewards, which the \textit{Critic} later uses. Finally, current and future state-action (and their corresponding rewards) are stored in the \textbf{Memory Module} for future reference either by the \textit{Actor} or the \textit{Critic}. 

Despite their clear functionality, several scientific and technical challenges remain in implementing the above-described foundation models. We discuss next these challenges and possible approaches for their development and implementation. 

\subsection{Implementing the Time-Series Foundation Model}\label{sec:time-series-foundation}

The primary task of the Time-Series Foundation Model is two-fold: first, to provide the State Estimator Foundation Model with the necessary time-series data to properly estimate the current system state; and second, to provide estimates (or forecasts) of exogenous variables, such as weather-dependent renewable generation and demand. The data used to train this model will be derived from different power system monitoring infrastructures (e.g., SCADA and PMU), which are inherently measured at different time scales (e.g., 1 s, 1 min, 5 min, 15 min, 1h). As a result, the expectation is that the Time-Series Foundation Model will seamlessly generate time-series data at any required time scale, helping fill in missing information, whether temporal or geographic. 

In generative AI, new sample-generated data that follows an empirical data distribution is known as \textit{synthetic data}. Our research showed that powerful generative AI models, such as Variational Autoencoders (VAEs) and Generative Adversarial Networks (GANs), are prone to collapse when used to generate synthetic time-series energy data~\cite{XiaHanyueSalazarShengrenPalensky2024}. Model collapse can lead to a failure to properly reproduce some time-series energy data features, thereby missing information relevant to power systems, such as demand peaks. This means that new deep learning architectures are needed to develop such a Time-Series Foundation Model. Specifically, such new architecture must be capable of
\begin{itemize}
    \item \textit{following the physics}; the generated synthetic data must follow the underlying power systems' physical phenomena. 
    \item \textit{adapting to multiple energy patterns}; a particular deep learning architecture may be enough to reproduce the time-correlation of one type of distribution (e.g., PV generation), while underperforming on others. Enough generalization capabilities for different energy demand patterns and their temporal and geographic correlations are a necessary feature. 
    \item \textit{modeling extreme values in the time-series data}; low-probability events, such as demand or renewable generation peaks, must be reproducible, as they play an important role in anticipating appropriate actions to account for them. 
    \item \textit{deploying conditioning features}; capable of generating data that respects the intrinsic patterns of the measured time-series data while satisfying either point (e.g., a peak value) or aggregated conditioned constraints (e.g., an average or total value). 
    \item \textit{adapting to multiple time-series scales to deploy down-sampling and up-sampling features}; generating data to recover high-frequency information from coarser-resolution measurements may be needed, ultimately reducing the need to update measurement infrastructure. 
    \item \textit{deploying data imputation}; filling in data missed, for instance, via loss of communication or faulty measurement units. 
\end{itemize}

Despite recent research efforts~\cite{NEURIPS2024_d0a2279c} (including ours, e.g., EnergyDiff~\cite{LinPalenskyVergara2025} and FCPFlow models~\cite{XiaWangPalenskyVergara2025}) to deploy the aforementioned features, a model that unifies all these features within a single ML architecture remains missing. This raises the question of \textit{how to develop a unified mathematical description of multiple generative tasks that can lead to a single ML architecture?.} A single ML architecture is needed to eliminate the need to retrain the Time-Series Foundation Model for different generation tasks and time resolutions, resulting in a more computational- and data-efficient approach. Recently, we validated the idea of such a unified generative task description, resulting in a single model for synthetic time-series data generation~\cite{LinYanboJaccoPalenskyVergara2026}. Nevertheless, further research in this direction is needed. Moreover, until now, most research has focused on enhancing the capabilities of generative AI algorithms to accurately represent temporal correlations among different energy patterns, while disregarding their geographic correlations. Research to consider both temporal and geographic correlations is still needed. 

\subsection{Implementing the Power System Foundation Model}\label{sec:power-syste-foundation}
The primary task of the Power System Foundation Model is to provide an estimation of the power systems future state based on high-fidelity simulations. This simulated information must be provided at different time scales using a single-model approach, enabling the \textit{Foundation Twin} to be used across different time horizons. This fundamental feature is expected to enable knowledge transfer (e.g., dynamics and uncertainty) across multiple power system decision-making processes (see Fig.~5 in~\cite{Zomerdijk2024}). The expectation is that a single-model approach will be more computational- and data-efficient than having separate power system models for different time scales. In general, the Power System Foundation Model must be capable of
\begin{itemize}
    \item \textit{following the physics}; representing accurately multiple power system physics phenomena (e.g., transients, steady-state).  
    \item \textit{making predictions across multiple timescales, time horizons, and geographic scopes}; resulting in a single model that is capable of simulating power systems phenomena usually described separately by ODEs and DAEs.
    \item \textit{being invariant towards the power system configuration}; that often varies due to several decision-making processes (e.g., network reconfiguration) or the natural system expansion due to new assets and resources being deployed.
    \item \textit{representing uncertainty}; power systems are not deterministic, and as a result, we must be able to account for errors and uncertainty.
\end{itemize}

Developing a single model that deploys these features is complex, and many of the current open challenges (e.g., informing physics, scalability, convergence, and uncertainty) in the ML scientific community also apply to such an envisioned model. I will explicitly focus on one of these challenges, namely, \textit{informing} the power systems physics into ML models. Currently, several approaches are available in the physics-informed ML community~\cite {Karniadakis2021}. Among these approaches, the most popular one informs power system physics via penalty terms in the loss function during deep learning model training. As a result, such deep learning models can better generalize and follow the physics, even when used for data extrapolation (i.e., predictions outside the domain of the training dataset). Nevertheless, it is not yet clear how such an approach can be applied to train deep learning architectures to enable multi-timescale modeling, raising questions such as \textit{how can we inform the physics at the right timescale without resulting in an untractable model?}. To answer this question, one would first need to define which physics should be informed at each timescale. This is important because incorporating all available physics via penalty terms may result in an overpenalized model that is not differentiable, fails to converge, or, ultimately, performs poorly. One possible approach to inform power systems' multi-timescale physics in ML models is to use a continuum representation (see, e.g.,~\cite{Henkes2022}), thereby avoiding the need for separate ODEs and DAEs. Nevertheless, power systems' continuum representation may not be applicable to all timescales, and this direction remains unexplored in large-scale power systems modeling. Another direction interesting to explore is the use of differentiable solvers and models (e.g., see the differentiable power flow formulation in~\cite{MuhammedDebus2026}). Nevertheless, it is also not clear how this can be generalized to cover all power systems' multi-timescale physics into a single ML architecture. 

\subsection{Implementing the State Estimator Foundation Model}\label{sec:state-estimation-foundation}

The primary task of the State Estimator Foundation Model is to estimate the current power system state at any time scale. Given its similar task to the Power System Foundation Model, similar features and implementation approach are expected. This includes using a single model to achieve computational and data efficiency, as well as enabling knowledge transfer across different time scales. In addition to the expected capabilities of the Power System Foundation Model described in Sec.~\ref{sec:power-syste-foundation}, the State Estimation Foundation Model must be capable of
\begin{itemize}
    \item \textit{handling state synchronization at multiple time scales}; providing synchronous state estimations for processes that operate at different time scales (see Fig.~\ref{fig:hierarchical_learning}).   
\end{itemize}

In general, multi-timescale state synchronization is an open challenge from the DTs' perspective, and interestingly, the expected features of the State Estimator Foundation Model can help to resolve it. Here, the State Estimator Foundation Model is framed as a simulation model (as explained before, similar to the Power System Foundation Model), and an open challenge in its development concerns state representation, specifically on \textit{how to properly and efficiently represent power systems' physical state in ML models?}. Different approaches are available to address this challenge: A long-vector representation has already been abandoned due to the high dimensionality of the resulting vector required to encode all the power system information, which can lead to inefficient learning. The length of such a vector representation increases even further if information from multiple timescales needs to be considered. Nevertheless, such a vector representation could make sense if a sequential data transformation is used, aiming to better exploit efficient sequential-data learning architectures such as the Decision Transformer~\cite{ChenLuRajeswaramLeeGroverIgo2021}. For this approach, the main challenge is how to make such a state vector invariant to missing state information. 

A different, and perhaps more sophisticated, approach is to leverage the graph-like structure of power systems, leading to the current state of the art, where graph-based ML models (e.g., graph neural networks, or GNNs) are used as the base state representation. Despite their clear advantage, as shown by our own research (e.g., PowerFlowNet~\cite{LinOrfanoudakisCardenasGiraldo2024}), GNNs still struggle to deliver models that are invariant to power system configuration, missing the expected generalization capabilities. Research on efficient power system state representations for ML models is still needed. An interesting approach to addressing the state representation challenge is to adopt a hybrid model that combines sequence and graph data. Nevertheless, independent of the approach used, this must fit with the way in which the power systems physics is informed (see Sec.~\ref{sec:power-syste-foundation}), which can increase the complexity of both State Estimator and Power System Foundation Models.

\subsection{Implementing the Optimizer Module}\label{sec:optimizer}

Lastly, the \textit{Optimizer Module}, whose primary task is to train, validate, and execute a policy to perform an optimization task defined by the \textit{Manager}. As mentioned before, the \textit{Optimizer Module} is composed of two main sub-models: an \textit{Actor} and a \textit{Critic}, and follows the classical actor-critic approach in RL. The primary task of the \textit{Actor} is to compute or propose possible actions based on the current power system state (playing the role of the optimizer and explorer), while the primary task of the \textit{Critic} is to provide an estimate of future (accumulated) rewards, based on the future state transitions provided by the Power Systems Foundation Model. In the current state of the art, both the Actor and the Critic are implemented using deep learning models. Several RL architectures are available to enhance the training (e.g., replay buffer) and convergence speed when developing such deep learning models, including using multiple critic networks (resulting in the well-known TD3 architecture). Similar RL architectures could be of benefit when implementing the \textit{Optimizer Module}. In general, the \textit{Optimizer Module} must be capable of:
\begin{itemize}
    \item \textit{providing actions across multiple timescales and horizons}; capable of handling multiple power systems decision-making processes with a single RL architecture.
    \item \textit{deploying multiple policies, if needed}; although I argue for a single multi-timescale policy, one per timescale or optimization task can also be deployed. The Manager Module would be responsible for selecting the appropriate policy for the right task (which could span multiple time scales and horizons).  
    \item \textit{learning by interacting with the Power System Foundation Models}; learning with a model falls within the area of model-based RL. Advances in this area can also be exploited when developing the \textit{Optimizer Module}.
\end{itemize}

\begin{figure}[t]
  \centering
  \includegraphics[width=\linewidth]{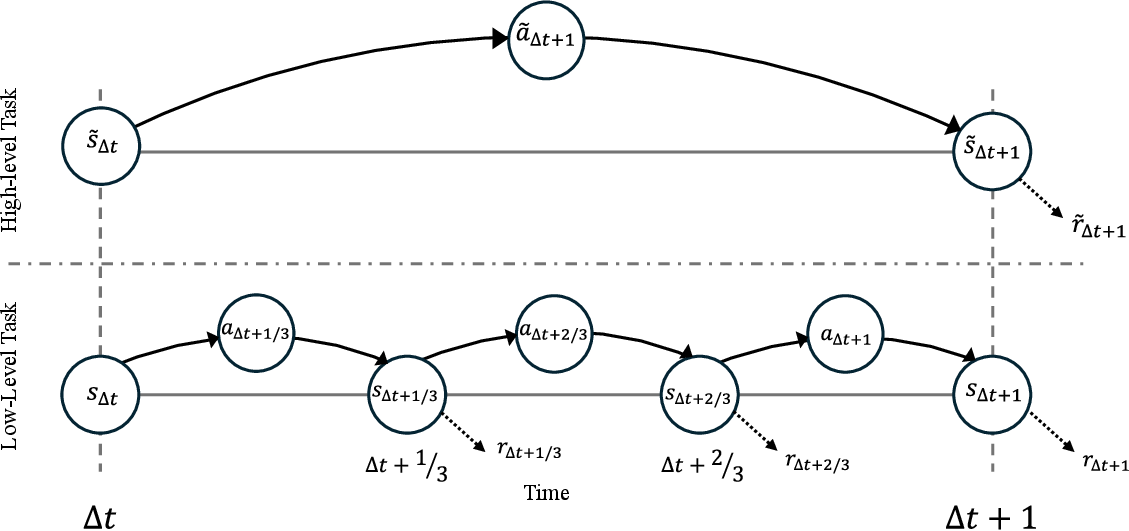}
  \caption{Representation of a hierarchical learning problem with two-level learning tasks: a low-level and a high-level. Here, actions must be taken at different time scales, with low-level actions also impacting the goal of the high-level task. An approach to coordinate both level tasks is to distribute the high-level goal task ($\tilde{r}_{\Delta +1}$) with the low-level sub-goals. For instance, enforcing $\tilde{r}_{\Delta t+ 1}=\sum^{3}_{i}r_{\Delta t +i/3}$. Note also the issue of state synchronization for both level tasks: At time $\Delta t$, state $s_{\Delta t} \approx \tilde{s}_{\Delta t}$. Similarly, at time $\Delta t+1$, state $s_{\Delta t +1} \approx \tilde{s}_{\Delta t +1}$. In power system applications, an example could be DERs dispatch in a market (high-level task) while accounting for voltage and frequency stability issues (low-level task).}
  \label{fig:hierarchical_learning}
\end{figure}

Providing decision-making capabilities across multiple timescales falls within the research area of Hierarchical RL. In Hierarchical RL, a long-term horizon (optimization) task is usually broken into sub-goals or commands that are easy to achieve, typically by learning a high-level controller that operates on longer timescales and provides low-level goals or commands to low-level controllers that provide short-term actions (see Fig.~\ref{fig:hierarchical_learning}). Nevertheless, hierarchical learning, which aims to deploy multi-timescale optimization capabilities, remains a largely unresolved problem in RL. In this regard, two main challenges are in place: first, \textit{how to design a feasible representation space for power system decision-making problems that fits into a hierarchical RL approach?}, and second, \textit{how to properly distribute long-term goals (from larger time scales) into smaller sub-goals (for shorter time scales)?}. The idea of designing a representation space, often called a latent space, to solve power system optimization tasks is to avoid using the original continuous space, where traditional RL architectures often struggle to converge during training. The approach proposed in~\cite{NEURIPS2022_a766f56d} has validated this idea for control of robots, while our SAVGO work has done so for other general optimization problems~\cite{OrfanoudakisVergara2026}. Nevertheless, transferring the optimization task to a new space may reduce its explainability and interpretability, features that must be retained in power system DTs. Moreover, it is unclear how such a latent space might look like for power system decision-making problems. Finally, breaking long-term goals into sub-goals to guide shorter-term optimization tasks aims to ease the task of achieving the long-term goals. Nevertheless, it is not clear how this can be done, especially if multiple timescales are to be considered. An interesting and promising approach to explore is the joint-embedded predictive architecture (JEPA), discussed in~\cite{Dawid2024}, which aims to coordinate decision-making across multiple timescales (reducing also the complexity of the decision-making problem). A different approach could be to deploy distinct policies to handle decision-making across different timescales. Nevertheless, this would result in separate models that may not be properly coordinated, or that cannot effectively share (fast and slow) dynamics, information, and decisions across all timescales. 

The above-mentioned RL challenges can be completely avoided if the proposed Optimizer Module is fully updated to use classical sequential decision-making approaches (e.g., model predictive control or stochastic programming). Such classical approaches deploy features needed for power systems optimization, including convergence guarantees and explainability, and have been widely studied. This suggests that an approach combining the fast decision-making and good generalization capabilities of RL algorithms with the convergence guarantees and explainability of classical approaches could be an interesting research direction. 

\subsection{Other Modules}
I have deliberately left out any discussion related to the \textit{Manager Module} and \textit{Memory Module}. I have done this because I consider these modules to be somewhat more trivial to develop than the foundation AI models that power the other components of the Foundation Twins architecture. As with any manager or coordination module, this must be developed with the proper human interface to enable the implementation of an Operator-in-the-Loop approach. Moreover, the Manager should ensure that other foundation AI models are properly updated when needed, using a continuous learning approach. Discussions of how to do this are important, but I consider them outside the scope of this paper. As per the \textit{Memory Module}, this must be able to store a large amount of state-action-reward data and, more importantly, retrieve it faster when required by the other foundation models.   

\section{Conclusion} \label{sec:conclusion}
This paper proposes a new generation of power system digital twins: Foundation Twins. Foundation Twins combine multiple foundation AI models with clearly defined capabilities and tasks that, together, aim to deliver the missing capabilities of current power system digital twins: multi-timescale decision-making in a single architecture that relies on high-fidelity simulation models. To do this, I presented the envisioned capabilities of each foundation AI model that comprises the proposed Foundation Twins architecture. To enable some of these capabilities, research in multiple directions is needed. I have also discussed several of these directions and the current open challenges that need to be addressed first, including some pressing ones: proper power system state representation in ML architectures and the effective incorporation of its multi-timescale physics. Finally, note that the ideas presented here are developed following a centralized approach, specifically, from the perspective of the network operators (Distribution and Transmission System Operators). Nevertheless, these ideas could also be extended to consider a distributed decision-making approach, perhaps by developing a Foundation Twin for each geographic scope, voltage level, etc. 

\section{Acknowledgment}
I would like to thank Dr. Juan S. Giraldo and Dr. Mauricio Salazar for (their friendship and) the time they spent checking this document and providing comments. I would also like to thank (our own soon-to-be Dr.) Stavros Orfanoudakis for his comments and ideas. 

\bibliographystyle{IEEEtran}
\bibliography{references}


\end{document}